\documentclass{article}
\usepackage{amsmath,epsfig}
\begin{document}


\begin{flushright}
\vspace{-10cm}
p. 121 in ``Topics in Condensed Matter Physics" Ed. M. P. Das \\ (Nova, New York 1994, ISBN 1560721804)
\end{flushright}

\vspace{3cm}
\centerline{{\bf QUASI-LOCAL}-{\bf DENSITY APPROXIMATION FOR A} }
                       
\centerline{{\bf VAN DER WAALS ENERGY FUNCTIONAL}}

\centerline{John F. Dobson }

\centerline{Faculty of Science and Technology, Griffith University, }

\centerline{Nathan, Queensland 4111, Australia}

\vspace{1cm}\centerline{{\bf ABSTRACT}}

\vspace{0.5cm}
We discuss a possible form for a theory akin to local density functional
theory, but able to produce van der Waals energies in a natural fashion. The
usual Local Density Approximation (LDA) for the exchange and correlation
energy $E_{xc}$ of an inhomogeneous electronic system can be derived by
making a quasilocal approximation for the {\it interacting} density-density
response function $\chi (\vec{r},\vec{r}\,^{\prime },\omega )$, then using
the fluctuation-dissipation theorem and a Feynman coupling-constant
integration to generate $E_{xc}$. The first new idea proposed here is to use
the same approach except that one makes a quasilocal approximation for the 
{\it bare} response $\chi ^{0}$, rather than for $\chi $. The interacting
response is then obtained by solving a nonlocal screening integral equation
in real space. If the nonlocal screening is done at the time-dependent
Hartree level, then the resulting energy is an approximation to the full
inhomogeneous RPA energy: we show here that the inhomogeneous RPA
correlation energy contains a van der Waals term for the case of
widely-separated neutral subsystems. The second new idea is to use a
particularly simple way of introducing LDA-like local field corrrections
into the screening equations, giving a theory which should remain reasonable
for all separations of a pair of subsystems, encompassing both the van der
Waals limit much as in RPA and the bonding limit much as in LDA theory. The
resulting functional is an explicit recipe which takes a trial electron
density distribution $n(r)$ as its input and yields $E_{xc}$ as its output.
The reason that it includes the van der Waals energy is that, like the RPA
formalism but unlike LDA, it involves spatially nonlocal screening for
inhomogeneous systems. It differs from earlier theories which effectively
use hydrodnamics to generate $\chi ^{0}$, in that rapid spatial variations
in electron density, such as those at subsystem boundaries, can be handled
more effectively; LDA-like local correlations are also included. Thus the
theory has credibility even at small separations outside the traditional
asymptotic van der Waals limit. We can therefore hope for a consistent
description of the forces between subsystems such as molecules or metal
surfaces, equally reasonable at all separations from the chemical bonding
regime through to the nonretarded van der Waals limit.

\vspace{1.5cm}\centerline{{\bf 1. INTRODUCTION}    }

\vspace{0.5cm}
Jay Mahanty is well known for his pioneering work on van der Waals (``vdW"
or ``dispersion") forces [1-7], and the present paper is presented as a
tribute in the form of a possible extension of his methods. Perhaps the most
familiar example of a dispersion force is the attractive interaction between
a widely separated pair of neutral atoms, which is proportional to $1/R^6$
in the nonretarded regime and to $1/R^7$ in the retarded regime, where $R$
is the interatomic separation. More generally, dispersion forces become
important in the interaction between electrically neutral subsystems at
separations larger than a few Angstroms so that electron transport between
the subsystems is negligible. The present paper attempts to generalise
existing theories of the van der Waals interaction to cases where the
separation is not necessarily large compared with atomic dimensions, so that
that the interaction between the subsystems cannot be treated
perturbatively. In this regime one has to consider Angstrom-scale details of
the electron distribution at the edges of the subsystems. For a description
of electronic behaviour on such short spatial scales one cannot rely upon
the hydrodynamic arguments which have recently been popular in the van der
Waals context, so a more microscopic theory is required. An additional
motivation for the present paper is the suggestion [4,8-11] that van der
Waals phenomena can be important even in large connected regions of high
electronic density $(e.g$. bulk metals) for which van der Waals phenomena
have not traditionally been considered.

The van der Waals energy is relevant in the interaction between neutral
systems in the domain where the separation is greater than a few Angstroms,
but less than the wavelength of light at the dominant fluctuation
frequencies so that electromagnetic retardation effects [1,3,5] can be
ignored. The vdW interaction can be derived, for example, by second-order
perturbation theory in the coulomb interaction between the two species [1],
or by an RPA-like fluctuation-dissipation theorem argument involving the
electric polarisability, which will be discussed further below. On the other
hand, for quantum systems in close contact, the simplest reasonably
successful general theory describing the Pauli repulsion and bonding forces
is the Local Density Functional theory of Kohn and Sham, together with
improvements by a number of workers [12,13]. The local density functional
expression for the exchange-correlation energy misses the van der Waals
interaction, however, and RPA calculations are not reliable for bonding
situations nor for extended systems at metallic densities. There is
therefore a need for a theory which works in all regimes. Ashcroft and
co-workers [8-11] have recently suggested ways of including van der Waals
effects in a density functional framework. In the present paper we suggest a
rather different quasi-local density formalism which should reproduce the
vdW force between widely separated species but which remains well-defined as
the species are brought together. It makes use of a prior calculation of the
dynamic electronic susceptibility of the uniform electron gas.

The new method will now be developed as follows. Firstly, in section 2 the
RPA groundstate energy formalism is reviewed for the case of an
inhomogeneous electronic system: this amounts to a time-dependent Hartree or
mean-field evaluation of the inhomogeneous dynamic susceptibility $\chi $
followed by use of the fluctuation-dissipation theorem in order to generate
approximate pair correlations. A coupling-constant integration is also
involved. Then in section 3 it is shown that the vdW interaction between
separated neutral systems is reproduced by the RPA formulation described in
Section 2. In Section 4 it is pointed out that the essential element in
producing the vdW energy term is not the nonlocality of the bare
susceptibility $\chi ^{0}$ (which merely reflects independent electron
transport, a phenomenon not involved in the vdW interaction). Rather, the
essence of the vdW interaction stems from nonlocality of the selfconsistent
Coulomb screening (Hartree) integral equation, which allows distant density
fluctuations to interact. It is therefore proposed to make a suitable local
or quasi-local approximation for $\chi ^{0}$, followed by solution of the
nonlocal Coulomb screening integral equation to produce an interacting
susceptibility $\chi $. Finally one uses the fluctuation-dissipation theorem
and a coupling-constant integration. This approach leads to an explicit
though fairly cumbersome procedure for obtaining a total energy including
the vdW contribution, starting from a given static density $n(r)$. This
expression is merely an approximation to the full RPA energy and does not
include the local field corrections inherent in, for example, the Kohn-Sham
exchange-correlation functional. To remedy this omission we propose in
section 5 a way to include LDA-like local field corrections alongside the
RPA effects already introduced. A summary is given in Section 6.

\vspace{1.5cm}\centerline{\bf 2. SUMMARY OF THE RPA GROUNDSTATE ENERGY
FORMALISM}

\vspace{0.5cm}
Since it is being asserted that the vdW interaction is inherent in the
inhomogeneous RPA correlation energy formalism, we begin by stating exactly
what is meant here by this formalism. To calculate the RPA linear response
of an interacting system one first solves the selfconsistent Hartree
groundstate equations [14] to yield a selfconsistent potential $V^{0h}(r,s)$%
, plus a set of one-particle orbitals $\varphi _{i}(r,s)$ and eigenvalues $%
\epsilon _{i}$. Then from first-order perturbation theory one forms [20] the
bare susceptibility

\begin{equation}
\chi ^{0}(\vec{r},\vec{r}\,^{\prime },\omega )=\sum_{i,j,s}(f_{i}-f_{j})\frac{%
\varphi _{i}^{*}(\vec{r},s)\varphi _{j}^{*}(\vec{r}\,^{\prime },s)\varphi
_{j}(\vec{r},s)\varphi _{i}(\vec{r}\,^{\prime },s)}{\epsilon _{i}-\epsilon
_{j}-\hbar \omega }.  \tag{1} 
\end{equation}
where $s$ labels the spin projection eigenvalue of a state. Physically, $%
\chi ^{0}$ represents the linear response, to an external potential $\delta
V(\vec{r})\exp (-i\omega t)$, of a system of {\it independent} electrons
moving in the static potential $V^{0h}(r):$i.e.

\begin{equation}
\delta n^{indep}(\vec{r},t)=\exp (-i\omega t)\int \chi ^{0}(\vec{r},\vec{r}%
\,^{\prime },\omega )\delta V(\vec{r}\,^{\prime })d^{3}r\,^{\prime }  \tag{2}
\end{equation}

The general form of (2) is dictated by the fact that it is the most general
possible linear time-invariant connection between the density perturbation $%
\delta n$ and the potential $\delta V$ which causes it. The fact that $\chi
^{0}$ is spatially nonlocal $(i.e$. connects different ``density'' and
``potential'' points $r$ and $r\,^{\prime })$ is solely due to the motion
(``transport'') of individual electrons between points $r$ and $r\,^{\prime }:$
it has nothing to do with electron-electron interactions since $\chi ^{0}$
refers to {\it independent} electrons. Only in the very simplest theories is 
$\chi ^{0}$ strictly local: for example static Thomas-Fermi theory assumes
slow spatial variations and takes $\chi ^{0}=\delta ^{3}(\vec{r}-\vec{r}%
\,^{\prime })\partial n(\vec{r})/\partial \mu (\vec{r})$. The derivation of
the exact nonlocal form (1) for $\chi ^{0}$ amounts to time-dependent
perturbation theory for each independent electron orbital $i$, followed by
squaring and summing over occupied states $i$ to obtain the density
perturbation at $r$. The energy denominator in (1) is the standard
denominator appearing in the wavefunction perturbation $\delta \psi _{i}$,
while the matrix elements $\delta V_{ij}$ of perturbation theory are
reproduced by the spatial integral in (2) acting on two of the wavefunctions
in (1). Alternatively (1) can be obtained from the real-space version of
Feynman diagram theory, being the retarded version of the
open-bubble-diagram polarisability $\pi ^{0}$ [16]. For the uniform electron
gas, (1) is simply the space Fourier transform of the familiar bare Lindhard
function.

While hydrodynamic approximations for $\chi ^{0}$ have the correct global or
large-distance behaviour, they do not obtain the correct $\vec{r}\simeq \vec{%
r}\,^{\prime }$ or high$-q$ behaviour embodied in (1). Thus hydrodynamic
approximations are less able to describe, for example, the detailed effects
of rapid electron-density falloff at a metal surface. In due course we will
introduce a quasi-local approximation for $\chi ^{0}$ which does retain the
correct short-ranged behaviour.

We turn now to the response of Coulomb-interacting electrons to an
externally imposed potential perturbation. The Time Dependent Hartree
Approximation [15], otherwise known as the Random Phase Approximation (RPA)
[16], amounts to the assumption that the electron density responds, via
equation (2), not to the external potential $\delta V^{ext}$, but rather to
a mean-field potential consisting of $\delta V^{ext}$ plus the Coulomb field
generated by the instantaneous density distribution itself. This plausible
``screening" argument entails neglect of the quantum or thermal fluctuations
of the density about its average value. Equivalently, it neglects the
correlations between the positions of the electrons: once an electron is
discovered at $r$, its coulomb repulsion reduces the probability of finding
other electrons nearby, so that the potential it feels due to the other
electrons is not precisely that generated by the mean density $n(r)$.

In general, for a sufficiently weak external potential perturbation \\$\delta
V^{ext}(r)\exp (-i\omega t)$, an interacting electron system must experience
a linear density response of form

\begin{equation}
\delta n(\vec{r},t)=\exp (-i\omega t)\int \chi (\vec{r},\vec{r}\,^{\prime
},\omega )\delta V^{ext}(\vec{r}\,^{\prime })d^{3}r\,^{\prime }.  \tag{3}
\end{equation}

The linear coefficient $\chi $ in the general case will be termed the
interacting susceptibility and can be regarded as the density response at $%
\vec{r}$ due to a potential perturbation localised at $\vec{r}\, ^{\prime }$.
The above mean-field argument then leads to the following Coulomb screening
equation for $\chi $ in the RPA:

\begin{equation}
\chi ^{rpa}(\lambda ,\vec{r},\vec{r}\,^{\prime },\omega ) = \chi ^{0}(\vec{r},%
\vec{r}\,^{\prime },\omega )  \nonumber 
\end{equation}

\vspace{-0.5cm}
\begin{equation}
{+\int \chi ^{0}(}\vec{r},\vec{r}_{1}{,\omega )\lambda e^{2}}|\vec{r}_{1}-%
\vec{r}_{2}|^{-1}\chi ^{rpa}(\lambda ,\vec{r}_{2},\vec{r}\,^{\prime },\omega
)d^{3}r_{1}d^{3}r_{2}{.}  \tag{4}
\end{equation}

Here the interaction strength parameter $\lambda $ is included for later
convenience and should for the present be set to unity.

The exchange and correlation energy $E_{xc}$ can now be obtained from the
interacting susceptibility $\chi $ in the following fashion. For a general
inhomogeneous electronic system, $E_{xc}$ can be expressed exactly in terms
of the density-fluctuation correlation function [17,18] as

\begin{equation}
E_{xc}=\frac{1}{2}\int_{0}^1 d\lambda \int d^{3}r\int d^{3}r\,^{\prime
}\ e^{2}|\vec{r}-\vec{r}\,^{\prime }|^{-1}\left\{ <\delta \hat{n}(\vec{r}%
)\delta \hat{n}(\vec{r}\,^{\prime })>_{\lambda }-n(\vec{r})\delta ^{3}(\vec{r}-%
\vec{r}\,^{\prime })\right\}  \tag{5}
\end{equation}
where $\delta \hat{n}(\vec{r})=\hat{n}(\vec{r},t)-n(\vec{r},t)$ is the
number-density fluctuation operator.

In the derivation of (5) it was assumed [17,18] that a $\lambda $-dependent
external potential $V(\lambda ,\vec{r})$ is supplied to maintain the density 
$n(\vec{r})$ at its fully-interacting value while the electron-electron
interaction strength is varied from $\lambda =0$ to $\lambda =1$. Without
the $\lambda $ integration, equ. (5) is clearly a contribution to the
potential energy arising from the correlations (both dynamical and
exchange-driven) between electron positions. The mutual avoidance of
electrons also leads to a kinetic energy contribution because of the
uncertainty principle, and qualitatively it is this feature which
necessitates the integration over the coupling-strength $\lambda $. Formally
the $\lambda $ integration is derived from a Feynman-theorem argument
[17,18].

We now use a frequency integral to generate the equal-time correlation
function $C(\vec{r},\vec{r}\,^{\prime },t=0)$ from its time Fourier
transform, and then apply the zero-temperature generalised
fluctuation-dissipation theorem [19], thus obtaining

\begin{equation}
< \delta \hat{n}(\vec{r})\delta \hat{n}(\vec{r}\,^{\prime })>_{\lambda
}=(2\pi )^{-1} \int_{-\infty}^{\infty }     C(\lambda ,\vec{r},\vec{r}%
\,^{\prime },\omega )d\omega  \nonumber 
\end{equation}

\begin{equation}
=-(\hbar /\pi ) Im\int_{0}^{\infty }d\omega \chi (\lambda ,\vec{r},%
\vec{r}\,^{\prime },\omega )  \nonumber
\end{equation}

\begin{equation}
=-(\hbar /\pi )\int_{0}^{\infty }ds\ \chi (\lambda ,\vec{r},\vec{r}%
\,^{\prime },is).  \tag{6}
\end{equation}

Here $\chi (\lambda ,\vec{r},\vec{r}\,^{\prime },\omega )$ is the exact linear
response of the interacting homogeneous system (with pair potential reduced
by a factor $\lambda )$ to an external potential $\delta V^{ext}(\vec{r}%
)\exp (-i\omega t)$. $\chi $ is in general defined as in equ. (3) except for
the presence of a reduced interaction strength:

\begin{equation}
\delta n_{\lambda }(\vec{r},t)=\exp (-i\omega t)\int \chi (\lambda ,\vec{r},%
\vec{r}\,^{\prime },\omega )\delta V^{ext}(\vec{r}\,^{\prime
})d^{3}r\,^{\prime }.  \tag{7}
\end{equation}

In (6) the frequency contour has been moved by analyticity arguments to lie
up the imaginary $\omega $ axis where $\chi $ is purely real. Putting
equations (5) and (6) together we have the following general exact result:

\begin{equation}
E_{xc} =\frac{1}{2}\int_{0}^1 d\lambda \int d^{3}rd^{3}r\,^{\prime }\
e^{2}|\vec{r}-\vec{r}\,^{\prime }|^{-1}  \nonumber 
\end{equation}

\vspace{-0.2cm}
\begin{equation}
\times \left\{ -\hbar \pi ^{-1}\int_{0}^{\infty }\chi (\lambda ,\vec{r},%
\vec{r}\,^{\prime },is)ds-n(\vec{r})\delta ^{3}(\vec{r}-\vec{r}\,^{\prime
})\right\}  \tag{8}
\end{equation}

This equation allows us to obtain the exchange-correlation energy of an
arbitrary inhomogeneous electronic system from a knowledge of the
interacting density-density response $\chi $. Essentially the same formula
was introduced by Harris and Jones [20] in order to investigate the surface
energy of a bounded jellium metal, the only difference being that in [20]
the density was not held constant by a $\lambda $-dependent external
potential, so that an extra electrostatic energy term was necessary.

In the case of independent electrons the exact susceptibility is the bare
susceptibility $\chi ^0$, and then, since there can be no dynamical
correlations, equation (8) gives the exact exchange energy $E_x$. The exact
correlation energy is then found [20] as the difference between the xc
energy and the exchange energy:

\begin{equation}
E_{c}=\frac{-\hbar }{2\pi }\int_{0}^1 d\lambda \int d^{3}r\int
d^{3}r\,^{\prime }\ e^{2}|\vec{r}-\vec{r}\,^{\prime }|^{-1}\int_{0}^{\infty
}(\chi (\lambda ,\vec{r},\vec{r}\,^{\prime },is)-\chi ^{0}(\vec{r},\vec{r}%
\,^{\prime },is))ds  \tag{9}
\end{equation}

In the present case we use the RPA approximation for $\chi $. Thus the RPA
groundstate energy of a general inhomogeneous system is defined as follows.
First we obtain the bare or independent-electron susceptibility from
equation (1), for the actual density $n(\vec{r})$. Then we obtain the
interacting susceptibility $\chi $ within the RPA by numerically solving the
screening integral equation (4), for each imaginary frequency $\omega =is$
and for each coupling strength from $\lambda =0$ to $\lambda =1$. Finally
the groundstate xc energy is obtained from (8), or equivalently the
correlation energy is obtained from (9).

The energy obtained in this fashion may be termed the RPA energy of the
inhomogeneous many-electron system at hand. This energy is a generalisation,
for inhomogeneous systems, of the familiar RPA (ring-diagram) groundstate
energy of the homogeneous electron gas [16]. It is worth stressing here that
this energy {\it does} include an approximate correlation energy, even
though, as discussed above, the time-dependent Hartree or RPA-screening
equations neglect the correlations between electron positions in the
presence of an external disturbance. The resolution of this apparent
contradiction lies in the application of the fluctuation-dissipation
theorem, which relates correlations of electron positions in the absence of
an external influence, to the response of electrons to the presence of a
time-varying external influence. The broken symmetry in the latter case
causes fluctuating density inhomogeneities to occur, and the mean-field
interactions between these inhomogeneities are the origin of the nontrivial
correlation function which emerges when the fluctuation-dissipation theorem
is applied to the RPA linear response function. This nontrivial correlation
function is in turn the origin of the van der Waals energy which we show
below is inherent in the RPA treatment of the groundstate energy of
inhomogeneous electronic systems.

Note that it is {\it not} being claimed that the RPA-screened Coulomb
interaction between two charges contains the vdW interaction as such: the
vdW interaction only emerges as a part of the correlation energy after use
of the fluctuation-dissipation theorem.

\vspace{1.5 cm}\centerline{\bf 3. VAN\ DER\ WAALS INTERACTIONS BETWEEN\
SEPARATED SYSTEMS}

\centerline{\bf \ FROM\ RPA\ CORRELATION ENERGY}

\vspace{0.5cm}
In order to show that the van der Waals interaction is inherent in the
inhomogeneous RPA correlation energy as described above, we now consider the
RPA description of the situation shown in Figure 1.

\centerline{\epsfig{figure=                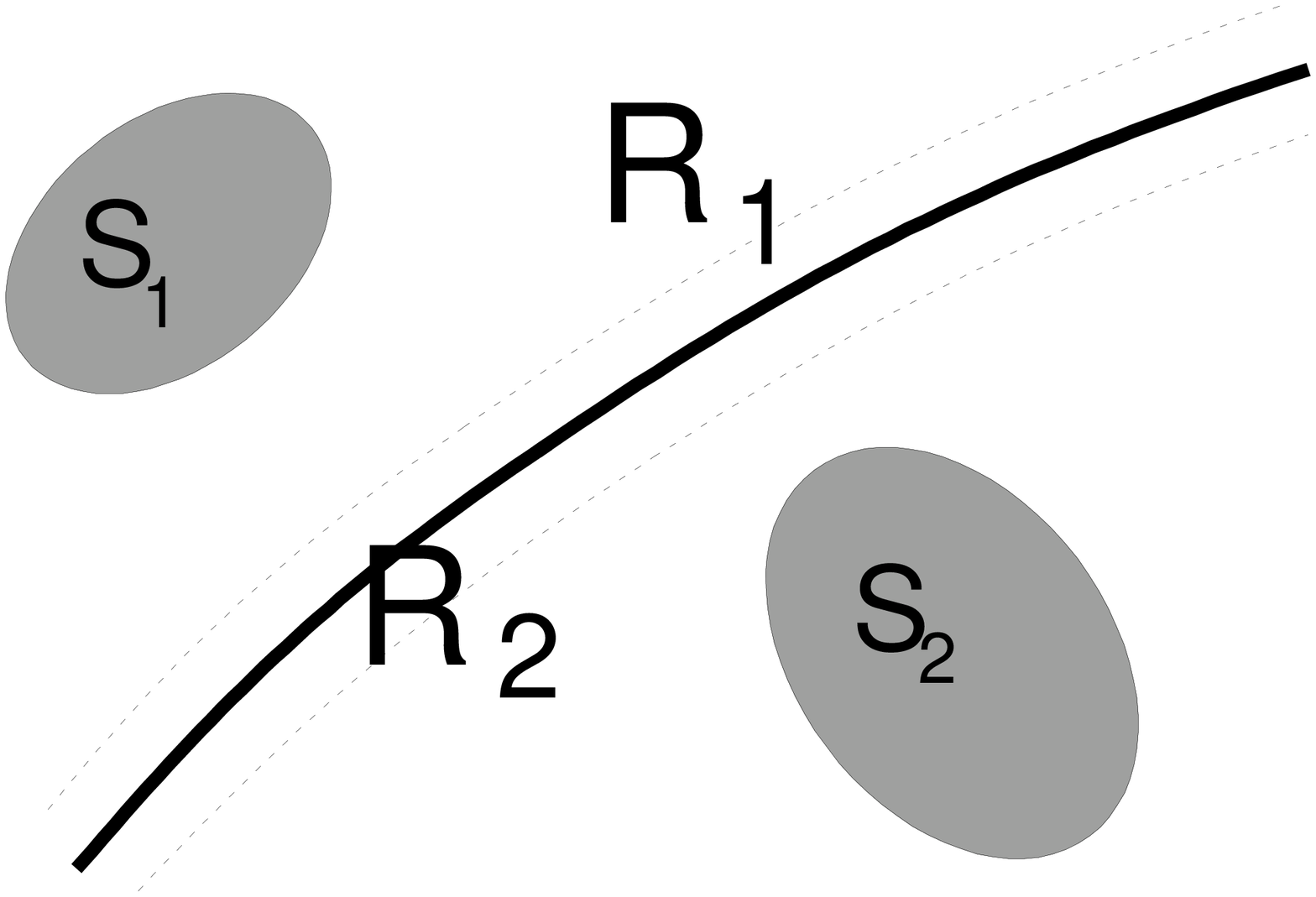,width=10cm}}

{\bf Figure 1}. Typical van der Waals situation. The heavy unbroken line
represents a fictitious surface dividing three-dimensional space into two
regions, ${\cal R}_1$ and ${\cal R}_2$. A localised charge-neutral
electronic system ${\cal S}_1$ lies inside ${\cal R}_1$ well away from the
boundary, and ${\cal S}_2$ is similarly located within ${\cal R}_2$. The van
der Waals part of the interaction between ${\cal S}_1$ and ${\cal S}_2$ is
present even if there is an impenetrable barrier between the dotted lines,
making electron transport between ${\cal R}_1$ and ${\cal R}_2$ impossible,
so that $\chi ^0{}_{12}$ is zero. In the general case the impenetrable
barrier is of course absent.


In Figure 1, two charge-neutral systems ${\cal S}_1$ and ${\cal S}_2$ are
separated by a large distance $R$ such that transport of electrons between $%
{\cal S}_1$ and ${\cal S}_2$ is highly improbable. We partition real space
into two regions ${\cal R}_1$ and ${\cal R}_2$ such that ${\cal R}_1$
contains ${\cal S}_1$ and ${\cal R}_2$ contains ${\cal S}_2$, with the
boundary between ${\cal R}_1$ and ${\cal R}_2$ well separated from both $%
{\cal S}_1$ and ${\cal S}_2.$

We now split the bare susceptibility $\chi ^0(r,r\,^{\prime},\omega )$ into
spatially partitioned form:


\begin{equation}
\chi ^{0}(\vec{r},\vec{r}\,^{\prime },\omega )=\chi _{1,1}^{0}(\vec{r},\vec{r}%
\,^{\prime },\omega )+\chi _{1,2}^{0}(\vec{r},\vec{r}\,^{\prime },\omega )+\chi
_{2,1}^{0}(\vec{r},\vec{r}\,^{\prime },\omega )+\chi _{2,2}^{0}(\vec{r},\vec{r}%
\,^{\prime },\omega )  \tag{10}
\end{equation}
where

\begin{equation}
\chi _{1,1}^{0}=\left\{ _{0\text{ otherwise}}^{\chi ^{0}(\vec{r},\vec{r}\,^{\prime
},\omega )\text{ if }\ \ \vec{r}\ \epsilon \ {\cal R}_{1}\text{ and }\vec{r}\,^{\prime
}\epsilon \ {\cal R}_{1}}\right\}  \tag{11a}
\end{equation}

\begin{equation}
\chi _{1,2}^{0}=\left\{ _{0\text{ otherwise}}^{\chi ^{0}(\vec{r},\vec{r}\,^{\prime
},\omega )\text{ if }\ \ \vec{r}\ \epsilon \ {\cal R}_{1}\text{ and }\vec{r}\,^{\prime
}\epsilon \ {\cal R}_{2}}\right\}  \tag{11b}
\end{equation}
and similarly for $\chi ^0{}_2,_1$ and $\chi ^0{}_2,_2.$

We seek to solve the screening equations in the case that the two neutral
systems are widely separated, so that $\chi ^0{}_1,_2$ (at the imaginary
frequencies we will be using) is exponentially small because of a tunneling
barrier. In the special case that there is an impenetrable barrier
separating ${\cal R}_1$ and ${\cal R}_2$, so that particle transport from $%
{\cal R}_1$ to ${\cal R}_2$ is impossible (see dotted lines in Fig. 1), then 
$\chi ^0{}_1,_2$ and $\chi ^0{}_2,_1$ can even be made exactly zero. This
special case emphasises the fact that the vdW energy we are about to obtain
does not depend on electron transport from ${\cal R}_1$ to ${\cal R}_2$, nor
on the presence of electron density in the region between ${\cal S}_1$ and $%
{\cal S}_2.$

As a reference consider first the case that ${\cal R}_1$ contains a neutral
species ${\cal S}_1$, but ${\cal R}_2$ is empty. (This case corresponds to
figure 1 with ${\cal S}_2$ removed). Then the screening equation (4) for the
interacting susceptibility $\chi ^{rpa}$, which we denote $\chi _1,_1$ in
this situation, becomes


\[
\chi _{1,1}(\vec{r},\vec{r}\,^{\prime },\omega )=\chi _{1,1}^{0}(\vec{r},\vec{r}\,^{\prime },\omega ) 
\]
\begin{equation}
+\int \chi _{1,1}^{0}(\vec{r},\vec{r}_{1},\omega )e^{2}|\vec{r}_{1}-\vec{r}_{2}|^{-1}\chi
_{1,1}(\vec{r}_{2},\vec{r}\,^{\prime },\omega )d^{3}r_{1}d^{3}r_{2}.  \tag{12}
\end{equation}

A similar equation holds for the case that ${\cal R}_{1}$ is empty, with
(1,1) replaced by (2,2) throughout.

Now consider the full RPA screening equation (4) for the van der Waals
problem, with both polarisable species ${\cal S}_1$ and ${\cal S}_2$ present
as in Figure 1. We write the combined interacting susceptibility as the sum
of the individual interacting susceptibilities plus a correction,

\begin{equation}
\chi =\chi _{1,1}+\chi _{2,2}+\Delta \chi .  \tag{13}
\end{equation}

We spatially partition the Coulomb interaction $V^c(\vec{r},\vec{r}\,^{\prime}) =
e^2|\vec{r}-\vec{r}\,^{\prime}|^{-1}$ and also the interacting susceptibility correction $%
\Delta \chi $, as in equation (11). Then, using * to represent spatial
convolution, we can write equation (4) as four separate equations, one for
each of the (1,1), (1,2), (2,1) and (2,2) cases. The first two of these
equations are

\begin{equation}
\chi _{1,1}+\Delta \chi _{1,1} =\chi _{1,1}^{0}+(\chi
_{1,1}^{0}*V_{1,1}^{c}+\chi _{1,2}^{0}*V_{2,1}^{c})*(\chi _{1,1}+\Delta \chi
_{1,1)}  \nonumber
\end{equation}

\vspace{-0.2cm}
\begin{equation}
+(\chi _{1,1}^{0}*V_{1,2}^{c}+\chi _{1,2}^{0}*V_{2,2}^{c})*\Delta \chi
_{2,1}  \tag{14}
\end{equation}

\begin{equation}
\Delta \chi _{1,2} =\chi _{1,2}^{0}+(\chi _{1,1}^{0}*V_{1,1}^{c}+\chi
_{1,2}^{0}*V_{2,1}^{c})*\Delta \chi _{1,2}  \nonumber
\end{equation}

\vspace{-0.2cm}
\begin{equation}
+(\chi _{1,1}^{0}*V_{1,2}^{c}+\chi _{1,2}^{0}*V_{2,2}^{c})*(\chi
_{2,2}+\Delta \chi _{2,2}).  \tag{15}
\end{equation}

Similar equations hold for $\Delta \chi _2,_1$ and $\Delta \chi _2,_2$. We
assume $\chi ^0{}_1,_2$ and $\chi ^0{}_2,_1$ are negligible, corresponding
to the vdW limit in which electron transport from ${\cal S}_1$ to ${\cal S}%
_2 $ is exponentially supressed. Quantities such as $\chi
^0{}_1,_1*V^c{}_1,_2$ will be treated as perturbations. This is reasonable
for well-separated subsystems since $\chi ^0{}_1,_1$ restricts $r$ to lie
deep inside region ${\cal R}_1$ while the index 1,2 on $V^c$ requires $%
\vec{r}\,^{\prime}$ to lie in region ${\cal R}_2$ so that $|\vec{r}-\vec{r}\,^{\prime}|$ is large,
making $V^c$ small. The interacting susceptibility correction $\Delta \chi $
is generated by switching on $(V^c{}_1,_2 + V^c{}_2,_1)$, and so all of its
partitioned components $\Delta \chi _1,_1$, $\Delta \chi _1,_2$ etc are of
first order in the perturbation, or smaller.

Keeping only zeroth and first-order terms in (15) we find


\begin{equation}
(I-\chi _{1,1}^{0}*V_{1,1}^{c})*\Delta \chi _{1,2}=\chi
_{1,1}^{0}*V_{1,2}^{c}*\chi _{2,2}+2ndorder.  \tag{16}
\end{equation}


Using


\begin{equation}
(1-\chi _{1,1}^{0}*V_{1,1}^{c})^{-1}*\chi _{1,1}^{0}=\chi _{1,1}  \tag{17}
\end{equation}
from equation (12), we can write (16) as

\begin{equation}
\Delta \chi _{1,2}=\chi _{1,1}*V_{1,2}^{c}*\chi _{2,2}+\{2nd\text{ order in }%
\ V_{1,2}^{c}\text{, }V_{2,1}^{c}\}.  \tag{18a}
\end{equation}
and similarly

\begin{equation}
\Delta \chi _{2,1}=\chi _{2,2}*V_{2,1}^{c}*\chi _{1,1}+\{2nd\text{ order in }%
\ V_{2,1}^{c}\text{, }V_{1,2}^{c}\}.  \tag{18b}
\end{equation}


Now expanding equation (14) to lowest nonvanishing order in $V^c{}_1,_2$ and 
$V^c{}_2,_1$, removing zeroth order terms by using (12) in the form \\$\chi
_1,_1 = \chi ^0{}_1,_1 + \chi ^0{}_1,_1*V^c{}_1,_1*\chi _1,_1$, and then
using (17) and (18b) we obtain


\begin{equation}
\Delta \chi _{1,1} =\chi _{1,1}*V_{1,2}^{c}*\Delta \chi
_{2,1}+\{higher\,\,order\}  \nonumber
\end{equation}

\vspace{-0.2cm}
\begin{equation}
=\chi _{1,1}*V_{1,2}^{c}*\chi _{2,2}*V_{2,1}^{c}*\chi _{1,1}  \nonumber
\end{equation}

\vspace{-0.2cm}
\begin{equation}
+\{3rd\,\,order\,\,in\,\,V_{1,2}^{c},V_{2,1}^{c}\}.  \tag{19}
\end{equation}

Equations (18) and (19), along with the (2,2) counterpart of (19), summarise
what happens to the RPA susceptibility when we switch on the Coulomb
interaction between two well-separated neutral subsystems. When used with
the fluctuation-dissipation theorem and a coupling-constant integration as
in (9), equations (18) and (19) give rise to a term in the correlation
energy which is precisely the vdW interaction. To see this, note that the
correlation energy expression (9) calls for the susceptibility $\chi $ at
reduced interaction strength $\lambda $. Thus in (18) and (19) (but not in
the explicit $|\vec{r}-\vec{r}\,^{\prime}|^{-1}$ factor appearing in (9)) we replace $%
V_1,_1$ by $\lambda V^c{}_1,_1$, $V^c{}_1,_2$ by $\lambda V^c{}_1,_2$, and
similarly for $V^c{}_2,_1$ and $V^c{}_2,_2$. Note that $\chi ^0$ is
independent of $\lambda $ and does not depend on $V^c$ (recall that there is
a single particle potential $V(\lambda ,\vec{r})$ holding the density constant).
Thus, the contribution to the correlation energy (9) which depends on $%
V^c{}_1,_2$ and $V^c{}_2,_1$ is


\begin{equation}
\Delta E_{c}=-\frac{\hbar }{2\pi }\int_{0}^{\infty }ds\int_{0}^{1}d\lambda \
A(\lambda ,s)  \tag{20}
\end{equation}
where
\begin{equation}
A(\lambda ,s) =\int \ d^{3}r^{\prime }\ V^{c}(\vec{r},\vec{r}\,^{\prime })\Delta \chi
(\lambda ,\vec{r},\vec{r}\,^{\prime },is)  \nonumber
\end{equation}

\vspace{-0.2cm}
\begin{equation}
=\int d^{3}r[V^{c}*\Delta \chi (\lambda ,is)]_{\vec{r},\vec{r}}=Tr[V^{c}*\Delta \chi
(\lambda ,is)]  \tag{21}
\end{equation}
where the trace is defined with respect to integration over $r$, and has the
usual cyclic property. Expanding with respect to the spatial partitioning
introduced earlier we have



\begin{equation}
A(\lambda ,s) =Tr[V_{1,1}^{c}*\Delta \chi _{1,1}(\lambda
,is)]+Tr[V_{1,2}^{c}*\Delta \chi _{2,1}(\lambda ,is)]  \nonumber
\end{equation}

\vspace{-0.2cm}
\begin{equation}
+Tr[V_{2,1}^{c}*\Delta \chi _{1,2}(\lambda ,is)]+Tr[V_{2,2}^{c}*\Delta
\chi _{2,2}(\lambda ,is)]  \nonumber
\end{equation}

\vspace{-0.5cm}
\begin{equation}
=A_{1,1}+A_{1,2}+A_{2,1}+A_{2,2}  \tag{22}
\end{equation}

Here, from (19) followed by permutation of the last factor to the front,

\begin{equation}
A_{1,1} =Tr[V_{1,1}^{c}*\chi _{1,1}(\lambda ,is)*\lambda V_{1,2}^{c}*\chi
_{2,2}(\lambda ,is)*\lambda V_{2,1}^{c}*\chi _{1,1}(\lambda ,is)]  \nonumber
\end{equation}

\vspace{-0.2cm}
\begin{equation}
=\lambda ^{2}Tr[\chi _{1,1}*V_{1,1}^{c}*\chi _{1,1}*V_{1,2}^{c}*\chi
_{2,2}*V_{2,1}^{c}]  \tag{23}
\end{equation}

It is interesting to note that, in order to obtain all terms in (22) which
are of second order in $V^c{}_1,_2$, we needed the second-order result in
(19) but only the first-order result in (18).

Now by differentiating the operator identity $\epsilon (\lambda
)^{-1}*\epsilon (\lambda ) = I$ we can show in general that


\begin{equation}
\frac{\partial }{\partial \lambda }[\epsilon (\lambda )^{-1}]=-\epsilon
(\lambda )^{-1}*\frac{\partial \epsilon }{\partial \lambda }*\epsilon
(\lambda )^{-1}  \tag{24}
\end{equation}
and applying this to the case $\epsilon _1,_1(\lambda ) = I - \lambda \chi
^0{}_1,_1*V^c{}_1,_1$ we find

\begin{equation}
\frac{\partial }{\partial \lambda }\chi _{1,1}(\lambda ,is) =\frac{%
\partial }{\partial \lambda }[\epsilon _{1,1}^{-1}*\chi _{1,1}^{0}] 
\nonumber
\end{equation}

\vspace{-0.2cm}
\begin{equation}
=-\epsilon _{1,1}^{-1}*(-\chi _{1,1}^{0}*V_{1,1}^{c})*\epsilon
_{1,1}^{-1}*\chi _{1,1}^{0}  \nonumber
\end{equation}

\vspace{-0.2cm}
\begin{equation}
=\chi _{1,1}*V_{1,1}^{c}*\chi _{1,1}  \tag{25}
\end{equation}
which reproduces the first three operators in (23). Putting (25) into (23)
and working similarly with $A_{2},_{2}$ we have


\begin{equation}
A_{1,1}+A_{2,2} =\lambda ^{2}Tr[\frac{\partial }{\partial \lambda }(\chi
_{1,1})*V_{1,2}^{c}*\chi _{2,2}*V_{2,1}^{c}++\chi _{1,1}*V_{1,2}^{c}*\frac{%
\partial }{\partial \lambda }(\chi _{2,2})*V_{2,1}^{c}]  \nonumber 
\end{equation}

\vspace{-0.2cm}
\begin{equation}
=\lambda ^{2}\frac{\partial }{\partial \lambda }Tr[V_{1,2}^{c}*\chi
_{2,2}*V_{2,1}^{c}*\chi _{1,1}]  \tag{26}
\end{equation}

But from (22) and (18b)

\begin{equation}
A_{1,2}=\lambda \ Tr[V_{1,2}^{c}*\chi _{2,2}*V_{2,1}^{c}*\chi
_{1,1}]=A_{2,1}.  \tag{27}
\end{equation}

Thus (22) becomes

\[
A(\lambda ,s) = (\lambda ^2 \frac{\partial }{\partial \lambda } + 2\lambda )
Tr[V^c_{1,2}*\chi _{2,2}*V^c_{2,1}*\chi _{1,1}] = \frac{\partial }{\partial
\lambda }(\lambda ^2Tr[\,\, ]). 
\]

Thus the $\lambda $ integration in (20) can be done analytically, giving

\begin{equation}
\Delta E^{c} =-\hbar (2\pi )^{-1}\int d^{3}r\ d^{3}r^{\prime }\
d^{3}r_{1}d^{3}r_{2}e^{2}|\vec{r}-\vec{r}\,^{\prime }|^{-1}e^{2}|\vec{r}_{1}-\vec{r}_{2}|^{-1} 
\nonumber 
\end{equation}

\vspace{-0.2cm}
 \begin{equation}
\times \int_{0}^{\infty }ds\ \chi _{1,1}(r,r_{1},is)\chi
_{2,2}(\vec{r}_{2},\vec{r}\,^{\prime },is).  \tag{28}
\end{equation}

Here the convolutions have now been written out in full, and the
susceptibilities are evaluated at the full interaction strength $\lambda =1$%
. Equation (28) is the expression obtained via perturbation theory by
Zaremba and Kohn [21] for the van der Waals energy between two arbitrary
polarisable systems. This shows that the van der Waals interaction is
contained naturally within the standard inhomogeneous RPA groundstate energy
formalism. Of course, although the form of (28) is correct, the vdW energy
produced by the present RPA formalism will involve the RPA susceptibilities $%
\chi _{1},_{1}$ and $\chi _{2},_{2}$ of the two separated systems, rather
than the exact susceptibilities.

As an example of the use of equation (28) consider two point polarisable
dipoles located at $\vec{r}_{01}$ and $\vec{r}_{02}$, with scalar polarisabilities $%
\alpha _1(\omega )$ and $\alpha _2(\omega )$. Under the application of an
external potential \\$\Phi (\vec{r})\exp (-i\omega t) = -e^{-1}\delta V^{ext}(\vec{r})
\exp (-i\omega t)$, the dipole moment at $r_{01}$ is proportional to the
local value of electric field:

\begin{equation}
\vec{p}_{1}=\alpha _{1}(\omega )(-\vec{\nabla} \Phi (\vec{r}_{01}))=\alpha (\omega
)e^{-1}\vec{\nabla} \delta V^{ext}(\vec{r}_{01}).  \tag{29}
\end{equation}

Since a dipole moment $p_1$ is produced by moving an electron through a
displacement $\vec{x} = -e^{-1}\vec{p}_1$, this dipole moment can be represented in the
weak-field regime by a delta-function-derivative electron number density
perturbation:

\begin{equation}
\delta n_{1}(\vec{r},t) =\delta ^{3}(\vec{r}-[\vec{r}_{01}+\vec{x}/2])-\delta ^{3}(\vec{r}-[\vec{r}_{01}-\vec{x}/2])
\nonumber
\end{equation}

\vspace{-0.2cm}
\begin{equation}
\simeq -\vec{x}.\vec{\nabla} \delta ^{3}(\vec{r}-\vec{r}_{01})=+e^{-1}\vec{p}_{1}.\vec{\nabla} \ \delta
^{3}(\vec{r}-\vec{r}_{01}).  \nonumber
\end{equation}

\vspace{-0.2cm}
\begin{equation}
=e^{-2}\alpha (\omega )\vec{\nabla} \delta V^{ext}(\vec{r}_{01}).\vec{\nabla} \delta
^{3}(\vec{r}-\vec{r}_{01})  \tag{30}
\end{equation}

We determine the susceptibility of the isolated system ${\cal S}_1$ by
demanding that the density perturbation (30) is reproduced by the
generalised response equation (3). Thus we find for a point polarisable
dipole at $r_{01}$

\begin{equation}
\chi _{1,1}(\vec{r},\vec{r}_{1},\omega )=-e^{-2}\alpha _{1}(\omega )\vec{\nabla} _{r}\delta
^{3}(\vec{r}-\vec{r}_{01}).\vec{\nabla} _{\vec{r}_{1}}\delta ^{3}(\vec{r}_{1}-\vec{r}_{01}).
\tag{31}
\end{equation}

Putting (31) (plus the equivalent for $\chi _2,_2(\vec{r}_2,\vec{r}\,^{\prime}))$ into
(28) we find, with Einstein summation convention on indices $j$ and $k,$

\begin{equation}
\Delta E_{c} =-\hbar (2\pi )^{-1}\int_{0}^{\infty }ds\ \alpha
_{1}(is)\alpha _{2}(is)\times  \nonumber
\end{equation}

\vspace{-0.2cm}
\begin{equation}
\partial _{j}\partial _{1j}\partial _{2k}\partial _{k}^{\prime
}|\vec{r}-\vec{r}\,^{\prime }|^{-1}|\vec{r}_{1}-\vec{r}_{2}|^{-1}\left| _{\vec{r}=\vec{r}_{2}=\vec{r}_{01}\text{, }%
\vec{r}_{1}=\vec{r}\,^{\prime }=\vec{r}_{02}}\right.  \nonumber
\end{equation}

\vspace{-0.2cm}
\begin{equation}
=\frac{-3\hbar }{\pi \ R^{6}}\int_{0}^{\infty }ds\ \alpha _{1}(is)\alpha
_{2}(is),\,\,\,\,\,\,\,\,\,\,\,\,\,\,\,R=|\vec{r}_{01}-\vec{r}_{02}|  \tag{32}
\end{equation}
which is exactly the van der Waals interaction (see equ 1.18 of ref [1]).

\vspace{1.5 cm}\centerline{\bf 4. QUASI-LOCAL APPROXIMATION FOR THE}
\centerline{\bf INDEPENDENT-ELECTRON SUSCEPTIBILITY $\chi^0$}

\vspace{0.5cm}
The standard local density approximation [12] for the exchange-correlation
energy can be obtained by making a quasi-local approximation for the {\it %
interacting} susceptibility $\chi $, as follows: 
\begin{equation}
\chi (\lambda ,\vec{r},\vec{r}\,^{\prime },\omega )\simeq \chi ^{LDA}(\lambda ,\vec{r},\vec{r}\,^{\prime
},\omega )=\chi ^{unif}(\lambda ,n=n(\vec{r}),|\vec{r}-\vec{r}\,^{\prime }|,\omega ).
\tag{33}
\end{equation}

Here $\chi ^{unif}(\lambda ,n,|\Delta \vec{r}|,\omega )$ is the susceptibility of
a {\it uniform} electron gas of number density $n$. Putting the (obviously
spatially unsymmetric) Ansatz (33) into (8) one readily obtains 
\begin{equation}
E_{xc}=\int d^{3}rn(\vec{r})\epsilon _{xc}(n(\vec{r}))  \tag{34}
\end{equation}
where 
\[
\epsilon _{xc}(n)=-N^{-1}(\hbar /2\pi )\int_{0}^{1}d\lambda \int d^{3}r\int
d^{3}r^{\prime }\ \lambda e^{2}|\vec{r}-\vec{r}\,^{\prime }|^{-1} 
\]
\begin{equation}
\times [\int_{0}^{\infty }\chi ^{unif}(\lambda ,n,|\vec{r}-\vec{r}\,^{\prime }|,is)ds-\pi
\hbar ^{-1}n\delta ^{3}(\vec{r}-\vec{r}\,^{\prime })]  \tag{35}
\end{equation}
is the exchange-correlation energy per particle of a uniform electron gas of
number density $n.$

An inspection of (34) shows that the LDF approximation cannot yield the van
der Waals energy of a pair of separated electron systems ${\cal S}_1$ and $%
{\cal S}_2$ as in section 3 above. Consider for example two density
concentrations separated in real space by a region of zero density as
imposed by an impenetrable barrier (see dotted lines in Fig. 1). Then the
exchange-correlation energy from equ. (34) is the sum of the energies of the
two isolated systems, there being no contribution depending on the
separation because of the local character of the density dependence in (34).

The essential idea proposed here is that we should use an Ansatz like (33),
not for the interacting susceptibility $\chi $, but for the {\it bare}
susceptibility $\chi ^{0}$. (It might be objected that screening is well
known to reduce the effective range of response functions so that one should
only make local approximations for interacting response: this is certainly
true for the response $F(\vec{r},\vec{r}\,^{\prime })$ of the electron density at $r$ to
an external point {\it charge} density at $\vec{r}\,^{\prime }$. However $\chi $,
although commonly termed the ``density-density response'', in fact describes
the density disturbance at $r$ due to a point {\it potential} disturbance at 
$\vec{r}\,^{\prime }$. In this context it is noteworthy that, via the Schrodinger
equation, the electron wavefunctions respond in a quasi-local manner to the
local values of the potential, and not directly to the charge density which
creates the potential. As a result, both $\chi $ and $\chi ^{0}$ are more
localised than $F$. $\chi $ is not more localised than $\chi ^{0}$ in
general, however, and in certain cases $\chi $ is {\it less} local than $%
\chi ^{0}:$ this occurs for example for real frequencies near a plasmon
excitation [22], or for widely separated charge concentrations as in the
classic van der Waals problem studied above. Specifically, for the system $%
{\cal S}_{1}+{\cal S}_{2}$ studied in section 3 above, the special case with
a hard wall between ${\cal R}_{1}$ and ${\cal R}_{2}$ has $\chi
^{0}{}_{1},_{2}=0$ (a kind of locality) but $\chi _{1},_{2}\neq 0$ (a kind
of nonlocality: see equ. (18)).

The essence of the present argument is then to presume a form of locality
for $\chi ^0$, allowing its values for an inhomogeneous system to be
approximated from a knowledge of the bare susceptibility of a uniform
electron gas. This avoids the need to calculate wavefunctions in the
inhomogeneous situation. The nonlocality of the RPA-screened susceptibility $%
\chi $ is maintained by solving an explicitly non-local screening integral
equation in real space.

The simplest way to attempt this approach, which is merely an approximation
to RPA, is to set

\begin{equation}
\chi ^{0}(\vec{r},\vec{r}\,^{\prime },\omega )\simeq \chi ^{0,\text{unif}%
}(n_{av}(\vec{r},\vec{r}\,^{\prime }),|\vec{r}-\vec{r}\,^{\prime }|,\omega ).
\tag{36}
\end{equation}
where
\begin{equation}
\chi ^{0,\text{unif}}(n,r,\omega )=(2\pi )^{-3}\int \chi ^{0L}(n,k,\omega
)\exp (i\vec{k}.\vec{r})d^{3}k  \tag{37}
\end{equation}
is the usual bare uniform-electron-gas Lindhard function (Feynman diagram
bubble integral), Fourier-transformed into real space. The simplest form for
the average density is an unsymmetrical one,

\begin{equation}
n_{av}(\vec{r},\vec{r}\,^{\prime })=n(\vec{r})  \tag{38}
\end{equation}
which does however have the advantage of preserving particle number as will
be discussed elsewhere. One could also use a hydrodynamic approximation for $%
\chi ^{0}$. This approach has been extensively investigated by Mahanty,
Summerside and others [1,2,3]. To implement the hydrodynamic approach in
full one has effectively to solve spatial differential equations to obtain $%
\chi ^{0}$. Simple analytic results can then be obtained which are accurate
for the form of the van der Waals interactions at large separation. The
present method aims to work even for small separations and so must be valid
both for slow and for rapid spatial variations, whereas hydrodynamics is
expected to be valid only in the limit of slow variations. As will be
discussed elsewhere, the present type of approximation correctly obtains
some of the short-ranged, high$-q$ response properties missed by
hydrodynamics, and also avoids the solution of spatial differential
equations in obtaining $\chi ^{0}.$

Note that equations (38), (36), (4) and (8) [or (9)], solved in that order,
constitute a path from a chosen trial electron density $n(\vec{r})$ to an exchange
and correlation energy $E_{xc}$ which we have argued will contain the van
der Waals interaction, unlike the usual LDF prescription. This occurs
because the nonlocal screening integral equation (4) is retained: the
precise details of the nonlocal behaviour of the true $\chi ^0$ which are
lost in the ansatz (36) are, as we have argued, unimportant in obtaining the
vdW energy. The advantage of the present approach over a full RPA xc energy
calculation is that one obtains the bare susceptibility approximately from
that of a uniform electron gas, without the need to find the inhomogeneous
one-electron wavefunctions required as in equ. (1) for a full RPA
calculation.

When the above prescription is applied to the uniform electron gas one
obtains, after a little algebra, the uniform-gas correlation energy as given
by diagrammatic perturbation theory in the RPA or ``ring-diagram"
approximation (see for example equation 12.23 of [16]). This serves to
emphasise that we are dealing with an RPA type of approximation, applied in
this case however to an inhomogeneous situation.

A potential problem with the above algorithm in the van der Waals context is
that the Ansatz (36), (38) somewhat overestimates the bare response $\chi
^0{}_1,_2$ of the density in one isolated electron concentration ${\cal S}_1$
to a disturbance $\delta V^{ext}$ occurring in ${\cal S}_2$. For example, if 
$\vec{r}\ \epsilon \ {\cal S}_1$ and $\vec{r}\,^{\prime}\ \epsilon \ {\cal S}_2$, (38) and
hence (36) is nonzero even when there is a hard wall between ${\cal S}_1$
and ${\cal S}_2$ so that the true $\chi ^0(\vec{r},\vec{r}\,^{\prime},\omega ) = \chi
^0{}_1,_2$ is strictly zero. This difficulty is inevitable when the average
density $n_{av}$ used in (38) depends only on $n(\vec{r})$. This is probably not a
serious difficulty, however, since the spurious response $\chi
^0{}_1,_2{}^{spur}$ (which affects results mainly via the first term on the
right-hand-side of equation (15)) is of Friedel form and hence is of order $%
|\vec{r}_1-\vec{r}_2|^{-3}$. This spurious term competes with the right-hand side of
equ. (16) which is however of order $|\vec{r}_1-\vec{r}_2|^{-2}$ as it is formally the
field due to a net-charge-neutral distribution, giving a dipole potential in
leading order. Thus under a local approximation for $\chi ^0$ the spurious
response induced is negligible for large separations. More sophisticated
quasi-local approximations, ensuring that $\chi ^0{}_1,_2$ is not
overestimated, will be discussed elsewhere.

\vspace{1.5cm}\centerline{{\bf 5. INCLUSION OF LDA-LIKE CORRELATIONS}}

\vspace{0.5cm}
The above theory is only an approximation to RPA, and RPA is known to yield
poor exchange and correlation energies for most condensed matter systems,
even though it does contain the van der Waals interaction in the case of
widely-separated neutral subsystems. Here we propose a simple approximate
way to remedy this by including an LDA-like local exchange and correlation
term which largely avoids overcounting the RPA correlations already included.

We achieve this by adding a local-field term $F_{xc}$ to the time-dependent
Hartree screening equation (4), which then becomes

\begin{equation}
\chi _{\lambda }=\chi ^{0}+\chi ^{0}*(V^{c}+F_{xc\lambda })*\chi _{\lambda }
\tag{39}
\end{equation}
\begin{equation}
F_{xc\lambda }(\vec{r},\vec{r}\,^{\prime })=\delta ^{3}(\vec{r}-\vec{r}\,^{\prime })\lambda
^{-1}f_{xc}(\lambda ^{-3}n(\vec{r}))  \tag{40}
\end{equation}
where one possible choice for $f_{xc}$ is given for $\lambda =1$ by
\begin{equation}
f_{xc}(n)=d^{2}(n\epsilon _{xc}(n))/dn^{2}.  \tag{41}
\end{equation}

Here $\epsilon _{xc}$ is the exchange-correlation energy per particle of a
uniform electron gas of number density $n$. The choice (41) for $f_{xc}$
makes (39), for $\lambda =1$, identical with the defining equation of the
Time Dependent Local Density Approximation [23, 24] which is widely used
[23, 24, 25] for the finite-frequency response of many-electron systems. To
generalise this to the case of a reduced Coulomb interaction $\lambda
e^{2}|\vec{r}-\vec{r}\,^{\prime }|^{-1}$ as in equation (40) we have used the following
scaling argument. In a uniform zero-temperature electron gas with Coulomb
coupling constant $\lambda e^{2}$, the relevant quantum length is $%
a_{b}{}^{*}=\lambda ^{-1}\hbar ^{2}/me^{2}$, the effective Bohr radius. The
energy scale is $H^{*}=\lambda ^{2}me^{4}/\hbar ^{2}$, the effective Hartree
unit. Thus, in a zero-temperature electron gas with density $n$ and reduced
Coulomb interaction $\lambda V^{c}$, the local field term $%
f_{xc}=d^{2}(n\epsilon _{xc})/dn^{2}$ is given by

\begin{equation}
\epsilon _{xc\lambda }(n)=\lambda ^{2}\epsilon (\lambda ^{-3}n)\text{, }%
f_{xc\lambda }(n)=\lambda ^{-1}f_{xc}(\lambda ^{-3}n).  \tag{42}
\end{equation}

To complete the energy functional, $\chi (\lambda ,\vec{r},\vec{r}\,^{\prime },\omega =is)$
from equ. (39) is then integrated with respect to $\lambda ,\omega ,r$ and $%
r^{\prime }$ as in equ. (9), to yield $E_{c}.$

This theory, when applied to the homogeneous electron gas, gives the
following expression for the corrrelation energy per electron:

\begin{equation}
\epsilon _{c}^{u}=-\hbar (2\pi )^{-4}n^{-1}\int_{0}^{1}d\lambda \int d^{3}q%
\frac{4\pi e^{2}}{q^{2}}\int_{0}^{\infty }ds[\chi (\lambda ,n,q,is)-\chi
^{0L}(n,q,is)]  \tag{43a}
\end{equation}
where
\begin{equation}
\chi (\lambda ,n,q,is)=\chi ^{0L}(n,q,is)\{1-[4\pi \lambda
^{2}e^{2}q^{-2}+\lambda ^{-1}f_{xc}(\lambda ^{-3}n)]\chi ^{0L}(n,q,is)\}^{-1}
\tag{43b}
\end{equation}

Here $\chi ^{0L}$ is the bare Lindhard susceptibility of the uniform
electron gas [16]. The quantity $\epsilon ^{u}{}_{c}$ from equs. (43) is not
necessarily the same as the correlation part of the uniform-gas energy $%
\epsilon _{xc}$ already used in the definition (41) of the static $f_{xc}$.
The basic reason for this is that we have ignored the $q$ dependence and $%
\omega $ dependence of $f_{xc}$, corresponding to the Time Dependent Local
Density Approximation. In principle if we used the $f_{xc}(\lambda ,q,\omega
)$ which made (43b) the exact dynamical susceptibility of the uniform gas,
then an equation similar to (43a) would of course give the exact $\epsilon
_{xc}$ for a uniform gas. We could even contemplate the use of a
frequency-dependent but local approximation for the inhomogeneous-gas
quantity $f_{xc}(\lambda ,\vec{r},\vec{r}\,^{\prime },\omega )$ (see for example the
approach of Gross, Kohn and Iwamoto [26,27]) in order to carry out the
present xc energy scheme. The exact $f_{xc}$ of a uniform gas is only
approximately known, however, unlike the energy $\epsilon _{xc}$ which is
known essentially exactly from Monte Carlo calculations. Probably a simpler
route, if we are to avoid the inconsistency between $\epsilon ^{u}{}_{xc}$
(equs. 43) and the input values of $\epsilon _{xc}$, is to find a $q-and\
\omega $-independent function $f_{xc}(n)$ (not equal to $d^{2}(n\epsilon
_{xc})/dn^{2})$ such that the known $(e.g$. Monte-Carlo-derived) $\epsilon
_{xc}(n)$ of the uniform gas is reproduced by equations (43). This $%
f_{xc}(n) $ function for the uniform gas could be found once and for all by
solving (43), regarded as a nonlinear integral equation for the function $%
f_{xc}$. This is then the ``best'' local frequency-independent $f_{xc}$ for
use in the present theory, in the sense that the exact uniform gas xc energy
will be reproduced by the theory. Clearly such a frequency-independent $%
f_{xc}$ would represent some sort of average over finite frequencies, unlike
equ (41) which can be shown [28] to yield the exact static response within
Kohn-Sham LDA theory. In initial tests and applications it may be simplest
to use the definition (41) directly without attempting to impose a
consistency condition on equ. (43), however.

\vspace{1.5cm}\centerline{{\bf 6. SUMMARY}}

\vspace{0.5cm}
We have suggested a theory which combines elements of the RPA (highly
nonlocal explicit Coulomb screening) and of local density functional theory
(quasilocal approximation for the BARE inhomogeneous electron gas
susceptibility, plus a local field correction). As a result, after
application of the fluctuation-dissipation and Feynman theorems, it produces
the van der Waals interaction in a natural fashion. The inputs to the theory
are a trial inhomogeneous groundstate density $n(r)$ and a uniform-gas
response quantity $f_{xc}(n)$ from equation (41) (or, preferably, from a
self-consistent solution of equs. (43) with $\epsilon ^u{}_{xc}= \epsilon
_{xc})$. Starting from the trial $n(r)$ we form the approximate bare
inhomogeneous response $\chi ^0$ from (38) and (36). This $\chi ^0$ then
determines an inhomogeneous $\chi $ via numerical solution of the spatially
inhomogeneous screening integral equation (39). The xc energy then follows
from spatial, frequency, and coupling-strength integrations as in equs. (8)
or (9). Equations (38), (36), (39), and (9), in that order, thus constitute
a path from $n(r)$ to a correlation energy $E_c:$ in this sense we have a
true density functional. The energy is obtained from $n(r)$ without
calculating Kohn-Sham-like orbitals. Advantages of the theory should include
the ability to calculate the force between two neutral systems, yielding the
van der Waals interaction at large separation and LDA-like results at small
separation, while remaining well-defined and physically reasonable at
intermediate separations. While our calculational procedure for the total
exchange-correlation energy is more complicated than the LDA algorithm, it
is certainly tractable in quasi-one-dimensional systems such as a pair of
three dimensional jellium metals with a vacuum gap separating their parallel
surfaces. (For this case the formalism can probably be carried out [20, 25]
with the full bare response from equ. (1), thus providing a test of the
local Ansatz (36) in the van der Waals context). For arbitrary
three-dimensional situations the linear screening integral equation (39) is
probably the time-limiting step, and since it can be expressed approximately
as a matrix inversion problem it is amenable to parallel computing
techniques.

One drawback of the present theory is that, for two small well-separated
neutral systems, although the basic $1/R^6$ separation dependence of the Van
der Waals interaction will be reproduced, the coefficient may be too large.
This is because the quasilocal electron-gas estimate of the individual
polarisabilities is likely to be an overestimate, being based data for a
uniform gas whose energy levels are closely spaced, in contrast to the
widely-spaced levels of small finite systems. This drawback should not apply
to large systems such as a juxtaposed pair of metal surfaces. In this latter
geometry, moreover, there already exist [29] efficient algorithms for
generating suitable tranforms of some of the necessary $q$-space uniform-gas
quantities.

\vspace{1.5cm}\centerline{{\bf REFERENCES}}

\vspace{0.5cm}
1. J. Mahanty and B.W. Ninham, {\it Dispersion Forces}, London: Academic
Press, 1976.

2. J. Mahanty and B.V. Paranjape, ``Van der Waals interaction between a
molecule and a metal surface: effect of spatial dispersion", {\it Sol}. {\it %
State Commun}. Vol. 24 No. 9, 1977, pp. 651-3.

3. P. Summerside and J. Mahanty, ``Retarded dispersion interaction between
metals", {\it Phys}. {\it Rev}. $B$, Vol. 19, No. 6, 1978, pp 2944-9.

4. J. Mahanty and R. Taylor, ``Van der Waals forces in metals", {\it Phys}. 
{\it Rev}. $B$, Vol. 17, No. 2, 1978, pp 554-9.

5. P. Summerside, ``Surface properties of metals", $Ph.D$ thesis, Australian
National University, Canberra, ACT, Australia, 1979.

6. J. Mahanty, ``Velocity dependence of the van der Waals force between
molecules", {\it J}. {\it Phys}. $B:$ {\it Atom}. {\it Molec}. {\it Phys}.,
Vol 13, No. 22, 1980, pp 4391-6.

7. J. Mahanty and B.V. Paranjape, ``Effect of plasmon dispersion on van der
Waals interaction of molecules near a metal surface", {\it Surf}. {\it Sci}%
., Vol. 202, No. 1-2, 1988, pp. 335-42.

8. A. Maggs and N.W. Ashcroft, ``Electronic fluctuation and cohesion in
metals ", {\it Phys}. {\it Rev}. {\it Lett}., Vol. 59, No. 1, 1987, pp.
113-6.

9. N.W. Ashcroft, ``Electronic fluctuation, the nature of interactions and
the structure of liquid metals", {\it Nuovo Cimento}, Vol. 12D, No. 4-5,
1990, pp 597-617.

10. N.W. Ashcroft, ``Electronic fluctuations and the van der Waals metal", 
{\it Phil}. {\it Trans}. {\it R}. {\it Soc}. {\it London}, Vol. 334, 1991,
pp. 407-23.

11. K. Rapcewicz and N.W. Ashcroft, ``Fluctuation attraction in condensed
matter: a nonlocal functional approach", {\it Phys}. {\it Rev}. $B$, Vol.
44, No. 8, 1991, pp 4032-5.

12. W. Kohn and L.J. Sham, ``Self-consistent equations including exchange
and correlation effects", {\it Phys}. {\it Rev}., Vol. 140, No. 4A, 1965,
pp. 1133-8.

13. R.O. Jones and O. Gunnarsson, ``The density functional formalism, its
applications and prospects", {\it Rev}. {\it Mod}. {\it Phys}., Vol. 61, No.
3, 1989, pp. 689-746.

14. A.S. Davydov, ``{\it Quantum Mechanics}" Vol. 1, Oxford:Pergamon Press,
1965.

15. D.M. Newns, ``Dielectric response of a semi-infinite degenerate electron
gas", {\it Phys}. {\it Rev}. $B$, Vol. 1, No. 8, 1970, pp. 3304-22.

16. A.L. Fetter and J.D. Walecka, ``Quantum theory of many-particle
systems", New York: McGraw-Hill 1971.

17. D.C. Langreth and J.P. Perdew, {\it Solid State Commun}., Vol. 17, 1978, 
$p$. 1425.

18. O. Gunnarsson and B.I. Lundqvist, ``Exchange and correlation in atoms,
molecules and solids by the spin-density-functional method", {\it Phys}. 
{\it Rev}. $B$, Vol. 13, No. 10, 1976, pp 4274-98.

19. L.D. Landau and E.M. Lifshitz, ``Statistical Physics", Reading, Mass.:
Addison-Wesley, 2nd. Ed., 1969 (sec. 127: the unsymmetrised version of the
theorem is required).

20. J. Harris and R.O. Jones, ``The surface energy of a bounded electron
gas", {\it J}.{\it Phys}. $F:$ {\it Metal Phys}., Vol. 4, No. 8, 1974, pp
1170-86.

21. E. Zaremba and W. Kohn, ``Van der Waals interaction between an atom and
a solid", {\it Phys}. {\it Rev}. $B$, Vol. 13, No. 6, 1976, pp 2270-2285.

22. J.F. Dobson and G.H. Harris, ``Microscopic electronic susceptibility of
the jellium half-space: a successful average-density {\it ansatz} for
complex frequency", {\it J}.{\it Phys}. $C:$ {\it Solid State Phys}., Vol.
20, No. 36, 1987, pp. 6127-36.

23. A. Zangwill and P. Soven, ``Density-functional approach to local-field
effects in finite systems: photoabsorption in the rare gases", {\it Phys}. 
{\it Rev}. {\it A}, Vol. 21, No. 5, 1980, pp. 1561-72.

24. A. Liebsch, ``Density functional approach to the dynamic response at
metal surfaces: van der Waals reference plane position and excitation of
electron-hole pairs", {\it J}. {\it Phys}. $C:$ {\it Solid State Phys}., Vol
19, No. 25, 1986, pp. 5025-47.

25. J.F. Dobson and G.H. Harris, ``An additional surface plasmon mode of a
bare jellium aluminium surface from self-consistent microscopic
calculations" {\it J}. {\it Phys}. $C:$ {\it Solid State Phys}., Vol. 21,
No. 21, 1988, pp. L729-L734.

26. E.K.U. Gross and W. Kohn, ``Local density-functional theory of
frequency-dependent linear response", {\it Phys}. {\it Rev}. {\it Lett}.,
Vol. 55, No. 26, 1985, pp 2850-52: erratum {\it ibid}., Vol. 57, No. 7, $p$.
923.

27. N. Iwamoto and E.K.U. Gross, ``Correlation effects on the
third-frequency-moment sum rule of electron liquids", {\it Phys}. {\it Rev}. 
$B$, Vol. 35, No. 6, 1987, pp. 3003-4.

28. J.F. Dobson and J.H. Rose, ``Surface properties of simple metals via
inhomogeneous linear electronic response: I. Theory", {\it J}. {\it Phys}. $%
C:$ {\it Solid State Phys}., Vol. 15, No. 36, pp. 7429-56, Appendix A2.

29. G.H. Harris, A.J. O'Connor and J.F. Dobson, ``Efficient calculation of
bulk jellium electronic susceptibilities for use in the theory of metal
surfaces", {\it J}. {\it Phys}. $C:$ {\it Solid State Phys}., Vol. 21, No.
1, 1988, pp. 107-117.

\end{document}